# Compiler Design for Legal Document TranslationIn Digital Government

Youssef Bassil

*LACSC – Lebanese Association for Computational Sciences*
*Registered under No. 957, 2011, Beirut, Lebanon*

**Abstract**

One of the main purposes of a computer is automation. In fact, automation is the technology by which a manual task is performed with minimum or zero human assistance. Over the years, automation has proved to reduce operation cost and maintenance time in addition to increase system productivity, reliability, and performance. Today, most computerized automation are done by a computer program which is a set of instructions executed from within the computer's memory by the computer central processing unit to control the computer's various operations. This paper proposes a compiler program that automates the validation and translation of input documents written in the Arabic language into XML output files that can be read by a computer. The input document is by nature unstructured and in plain-text as it is written by people manually; while, the generated output is a structured machine-readable XML file. The proposed compiler program is actually a part of a bigger project related to digital government and is meant to automate the processing and archiving of juridical data and documents. In essence, the proposed compiler program is composed of a scanner, a parser, and a code generator. Experiments showed that such automation practices could prove to be a starting point for a future digital government platform for the Lebanese government. As further research, other types of juridical documents are to be investigated, mainly those that require error detection and correction.

## I. INTRODUCTION

Before the advent of computers, people used to do manual work without any electronic assistance. With time, operations became so huge so that it had to be administered constantly by humans, leading to a bureaucracy system that was laborious, time-consuming, prone to errors, and very expensive. As a result, the situation became epidemic and something had to be done. The answer turned out to be using the computer to automate manual processes and tasks [1]. At first, the word "mechanized" became common not until it was substituted by the more accurate word "automation". In fact, automated computer operations began in the late 1960's when IBM introduced the groundbreaking line of its mini computers and operating systems. One of the earliest successful models introduced by IBM was the OS/360 which was a supervisory batch software that controls system resources and delivers automatic transition from one task to another [2]. Thus the name batch processing was coined. The OS/360 compared to modern OS was a simple primitive task scheduler that could run limited batch jobs in sequence without any capability of decision making whatsoever. It was kind of a semi-automated system that still required human labor involvement in a sense that it had to be manually configured and managed by the operations staff. After decades of requiring humans to carry out labor-intensive tasks, software companies began developing real and fully automated operation software [3]. Many products were introduced to the market including but not limited to intelligent and complicated job scheduling, database management software, recovery systems, backup services, and computer aided design. This has led to a drastic reduction in cost and an increase in productivity, quality, and performance [4]. Basically, the backbone of every computer operation automation is the software or computer program. From a technical point of view, a computer program is a set of electronic instructions executed from within the computer's memory by the computer central processing unit CPU [5]. The purpose of a computer program is to control the functionalities of the computer allowing it to perform miscellaneous tasks ranging from mathematical computations to scientific operations, accounting, data management, gaming, text editing, audio, video, and image archiving, and Internet.

This paper proposes an automated software with the purpose of automating the conversion of unstructured plain-text documents into structured XML files [6]. The documents contain legal content mainly written in the Arabic language related to juridical proceedings, decree laws, and legal procedures. The core of the proposed software is a compiler system, made up of a scanner, a parser, and a code generator. The scanner is a finite automaton that reads and interprets input textual data; the parser is a context-free grammar device that validates the structure of the input; and the code generator is an output engine that produces XML construct code.

## II. BACKGROUND

Fundamentally, a compiler is a computer program that transforms a code written in a source language into another code written in a target language. Compilers are a type of automated translators that convert instructions and data from a high-level form into a machine-code form that can be read, processed, and executed by a computer.

A compiler consists of five major building blocks: The Preprocessor, the Scanner, the Parser, the Semantic Analyzer, and the Code Generator [7].

- *The Preprocessor*: Its purpose is to reduce the complexity of the input code to make the job easier on the scanner. The preprocessor has many tasks including removing annotations, getting rid of extra white-lines and white-spaces, and deleting unused symbols and notations.
- *The Scanner*: Its purpose is to tokenize the input code and divide it into meaningful tokens such as keywords and data values. The algorithm of the scanner is built upon Finite-State machine (DFA) [8] and Regular Expressions.
- *The Parser*: Its purpose is to detect syntax errors by performing Syntax Checking against the tokens generated by the scanner. The output is a Parse-Tree known as Syntax-Tree. Syntax Checking is about verifying that the arrangement of tokens as received from the scanner are in the correct order and comply with the grammar of the source language.
- *The Semantic Analyzer*: Its purpose is to perform Semantic Checking which consists of verifying that the written code comply with the semantics of the source language. Semantics are the different rules that define restrictions on the syntax.
- *The Code Generator*: Its purpose is to convert the parse-tree generated by the parser into a target code. The target code can be either Assembly code, Machine code, Bytes code, or even another high-level language such as XML or JSON.

## III. THE PROPOSED SOLUTION

This paper proposes a compiler software that automates the validation and translation of an input document written in a high-level language, namely the Arabic language, into an output document that can be read by computers. The input document is an Arabic unstructured plain-text originally written manually by people. On the other hand, the generated output document is a structured XML-based file that can be manipulated by computers. The proposed compiler is actually a part of a bigger project related to digital government and it is meant to automate the processing and archiving of juridical data and documents. In essence, the proposed compiler system is composed of a scanner, a parser, and a code generator. Figure 1 depicts the major components of the proposed compiler.

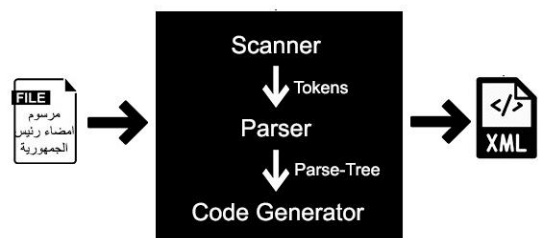

**Fig 1: Proposed Compiler System**

### A. The Scanner

The scanner is mostly based on Regular Expressions and DFAs short for Deterministic Finite-State Automata. Figure 2 and Figure 3 are samples of Finite Automata that the scanner uses to detect and tokenize numeric and string values respectively.

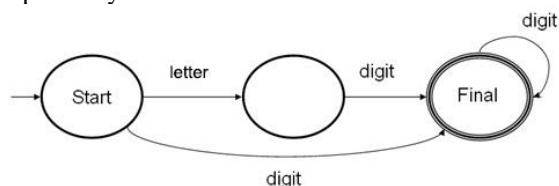

**Fig 2: Finite Automata for NUMERIC type values**

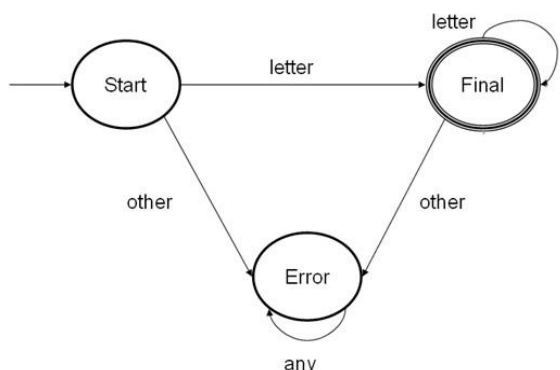

**Fig 3: Finite Automata for STRING type values**

Following are the regular expressions that are used by the scanner to match pattern of characters in the input file:

NUM = letter* digit digit*
STRING = letter letter*
digit = 0 | .. | 9
letter = ( أ | .. | ي ) | ( 0 | .. | 9 ) | ٠ | . | - | : | / | 4
TYPE = مرسوم | قرار | قانون
RAQM = رقم
INNA = إن
BINAA = ونظرا | وبعد الاطلاع | وبعد موافقة | وبناء على | بناء على
HAYSOU = نظرا | وبعد أن | وبما أن | وحيث أن
YAKOUR = يرسم ما يأتي | يرسم ما يلي | يقرر ما يأتي | يقرر ما يلي
NADA = مادة | المادة
FI = في
IMDAA = إمضاء | الإمضاء

### B. The Parser

The parser is built upon a formal grammar. It is based on a CFG or Context-Free Grammar [9] as it provides powerful features including but not limited to recursion, cascading, and nesting. Below is the CFG of the proposed parser:

document → statement title issuer ref-list just-list acknowledge article-list loc-date sig-list
statement→ TYPE RAQM NUM
title→STRING
issuer→INNA STRING ،
ref-list→ ref ref-list | ref
ref→ BINNA STRING ، | BINNA STRING .
just-list→ just just-list | λ
just→ HAYSOU STRING ، | HAYSOU STRING .
acknowledge→ YAKOUR **:**
article-list→ article article-list | article
article→ MADA article-num**:** article-title article-content
article-num→ NUM | STRING
article-title→ STRING | λ
article-content→ STRING
loc-date→ STRING FI STRING | STRING STRING
sig-list→ sig-type1 | sig-type2-list
sig-type1→ IMDAA **:** STRING STRING | λ
sig-type2-list→ sig-type2 sig-type2-list | sig-type2
sig-type2→ STRING IMDAA **:** STRING

### C. The Code Generator

The code generator is the final component of the compiler whose task is to generate the final XML document based on the parse-tree received from the parser. Below are the generation rules of the proposed code generator:

*Synthesized Attributes:*
- *s used for static source code*
- *c represents the content of XML tags*

*Inherited Attributes:*
- *s' symmetric to s*

*Rule 1:* document → statement title issuer ref-list just-list acknowledge article-list loc-date sig-list

s'(document) = '' using system; using system.IO; class program {static void main ( ) {string XML output = \ " "
s(document) = '' \"; File. Write All Text (@ \" c \\", XML output); }}"
c(document) = c(statement) U c(title) U c(issuer) U c(ref-list) U c(just-list) U c(acknowledge) U c(article-list) U c(loc-date) U c(sign-list)

*Rule 2:* statement → TYPE RAQM NUM

s' (TYPE) = s' (statement) = "<type>"
c (TYPE) = " \" + TYPE + \" "
s (TYPE) = "</type>"
s' (RAQM) = "<content Number>"
c (RAQM) = "\" + NUM + \"
s (RAQM) = "</content Number>"
c (statement) = c (RAQM) U c (TYPE)
s (statement) = s (RAQM)

*Rule 3:* title → STRING

s' (title) = "<title>"
c (title) = c (STRING) = " \ " + STRING + \" "
s (title) = "</title>"

*Rule 4:* issuer → INNA STRING ،

s' (issuer) = "<issuer>"
c (issuer) = c (STRING) = " \ " + STRING + \" "
s (issuer) = "</issuer>"

*Rule 5:* ref-list → ref ref-list

s' (ref) = s' (ref-list0)
s (ref-list1) = s (ref-list0)
c (ref-list0) = c (ref) U c (ref-list 1)
s' (ref) = s' (ref-list0) = "<references>"
s (ref-list0) = s (ref-list1) = "</references>"

ref-list→ ref

s' (ref) = s' (ref-list)
s (ref-list) = c (ref)
s (ref-list) = s (ref)
s' (ref) = s' (ref-list) = "<references>"
s (ref-list) = s (ref) = "</references>"

*Rule 6:* ref → BINAA STRING ،

s' (ref) = "<references>"
s (ref) = "</references>"
c (ref) = c (STRING) = "\" + STRING + \" "

ref→ BINAA STRING .

s' (ref) = "<references>"
s (ref) = "</references>"
c (ref) = c (STRING) = "\" + STRING + \" "

*Rule 7:* just-list →just just-list

S' (just) = S' (just-list0) = "<justifications>"
S (just-list0) = S (just-list1) = "</justifications>"
s' (just) = s' (just-list0)
s (just-list1) = s (just-list0)
c (just-list0) = c (just) U c (just-list1)

*Rule 8:* just → HAYSOU STRING ،

s' (just) = "<justifications>"
s (just) = "</justifications>"
c (just) = c (STRING) = " \ " + STRING + \" "

*Rule 9:* acknowledge → YAKOUR :

s' (acknowledge) = " "
c (acknowledge) = " "
s (acknowledge) = " "

*Rule 10:* article-list → article article –list

S' (article) = S' (article-list0) = "<articles>"
S (article-list0) = S (article-list1) = "</articles>"
s' (article) = s' (article-list0)
s (article-list1) = s (article-list0)
c (article-list0) = c (article) U c (article-list1)

*Rule 11:* article → MADA article-num : article-title article-content

S' (MADA) = s' (article) = "<articles>"
s (article) = S (MADA) = "</articles>"
c (article) = c (article-num) U c (article-title) U c (article-content)

*Rule 12:* article-num→ NUM

s' (article-num) = "< article Number >"

s (article-num) = "< /articles Number >"
c (article-num) = c (NUM) = "\" + NUM + \" "

*Rule 13*:  article-title → STRING

s' (article-title) = "<articleTitle>"
s (article-title) = "</articleTitle>"
c (article-title) = c (STRING) = "\" + STRING + \" "

*Rule 14*:  article-content → STRING

s' (article-content) = "<articleContent>"
s (article-content) = "</articleContent>"
c (article-content) = "\" + STRING + \" " = c (STRING)

*Rule 15*:  article-title → STRING

s' (article-num) = "<articleNumber>"
s (article-num) = "</articleNumber>"
c (article-num) = c (STRING) = "\" + STRING + \" "

*Rule 16*:  article-title → λ

s' (article-title) = " "
s (article-title) = "<articleTitle/>"
c (article-title) = c (λ) = " "

*Rule 17*:loc-date →STRING FI STRING

s' (STRING1) = s' (loc-date) = "<issueLocation>"
c (STRING1) = "\" + STRING 1 + \" "
s (STRING1) = "</issueLocation>"
s' (STRING2) = "<issueDate>"
c (STRING2) = "\" + STRING2 + \" "
s (loc-date) = s (STRING2) = "</issue Date>"
c (loc-date) = c (STRING1) U c (STRING2)

*Rule 18*:  sig-list → sigType1

S' (sigType1) = S' (sig-list) = "<signatures>"
s' (sigType1) = s' (sig-list)
c (sig-list) = c (sig Type1)
s (sig-list) = s (sig Type1)
S (sig-list) = S (sig Type1) = "</signatures>"

sig-list→ sigType2-list

S' (sigType2-list) = S' (sig-list) = "<signatures>"
s' (sigType2-list) = s' (sig-list)
c (sig-list2-list) = c (sig-list)
s (sig-list) = s (sigType2-list)
S (sig-list) = S (sigType2-list) = "</signatures>"

*Rule 19*:  sigType1 →IMDAA : STRING  STRING

s' (IMDAA) = s' (sigType1) = "<signature>"
s' (STRING1) = "<name>"
c (STRING1) = "\" + STRING 1 + \" "
s (STRING1) = "</name>"
s' (STRING2) = "<position>"
c (STRING2) = "\" + STRING2 + \" "
s (STRING2) = "</position>"
s (sigType1) = s (IMDAA) = "</signature>"
c (sigType1) = c (STRING1) U c (STRING2)

*Rule 20*:  sigType2-list →sigType2  sigType2-list

s' (sigType2) = s' (sigType2-list)
s' (sigType2-list0) = s' (sigType2-list1)
c (sigType2-list0) = c (sigType2) U c (sigType2-list1)

sigType2-list→ sigType2

s' (sigType2) = s' (sigType2-list)
c (sigType2-list) = c (sigType2)
s (sigType2-list) = s (sigType2)

*Rule 21*:sigType 2 →STRING  IMDAA : STRING

s' (STRING1) = s' (sigType2) = "<signature>"
s' (STRING1) = "<position>"
c (STRING1) = "\" + STRING1 + \""
s (STRING1) = "</position>"
s' (STRING2) = "<name>"
c (STRING2) = "\" + STRING2 + \""
s (STRING2) = "</name>"
s (sigType2) = "</signature>"

## IV.  EXPERIMENTS & CONCLUSIONS

In the experiments, a sample unstructured input document is fed to the developed compiler system. It is mainly a decree issued in the Arabic language that dates back to year 2018. Figure 4 depicts the input document; while, Figure 5 shows the output code generated by the compiler after translating the input document. Obviously, the output is a structured XML file conveying the data originally presented in the input file. All in all, the proposed system delivers automation capabilities for converting hand-written juridical files into digital XML documents that can be managed by modern computers. As the original decree file became in digital format, it can now be stored and archived on computers. Moreover, being structured, it can be easily and systematically digitally manipulated including searching, editing, extending, and printing. This automation could prove to be a starting point for a future digital government platform for the Lebanese government.

مرسوم رقم ٢٥
دعوة مجلس النواب إلى عقد استثنائي
إن رئيس الجمهورية،
بناء على الدستور لا سيما المادتان ٣٣ و ٨٦ منه،
بناء على اقتراح رئيس مجلس الوزراء،
يرسم ما يأتي:
مادة ١ : عقد استثنائي
يدعى مجلس النواب إلى عقد استثنائي يفتتح بتاريخ ١٩/٣/٢٠١٨
ويختتم بتاريخ ٢٠١٨/٥/١٩
مادة ٢: برنامج أعمال
يحدد برنامج أعمال هذا العقد الاستثنائي بما يلي:
- مشاريع موازنات الأعوام ٢٠١٦-٢٠١٧ و ٢٠١٨
- مشاريع القوانين والاقتراحات والنصوص التي يقرر مكتب المجلس طرحها على المجلس.
مادة ٣ :
ينشر هذا المرسوم ويبلغ حيث تدعو الحاجة
بعيدا في ١٣ آذار ٢٠١٨
الامضاء: ميشال عون
صدر عن رئيس الجمهورية
رئيس مجلس الوزراء
الامضاء: سعد الدين الحريري

**Fig 4: Input Document**

```xml
<?xml version = " 1.0" encoding = "UTF-8 ?>
<NewDataSet>
  <Table>
    <Type>مرسوم</Type>
    <ContentNumber>25</ContentNumber>
    <Title>دعوة مجلس النواب إلى عقد استثنائي</Title>
    <justifications>
      <Reference>
        إن رئيس الجمهورية، بناء على الدستور لا سيما المادتان
        ٣٣ و ٨٦ منه، بناء على اقتراح رئيس مجلس الوزراء،
        يرسم ما يأتي
      </Reference>
      </D_ContentDate>١٣ اذار ٢٠١٨<D_ContentDate>
      <Signature>
        بعدما في ١٣ اذار ٢٠١٨ الإمضاء: مثبال عون، صدر عن
        رئيس الجمهورية رئيس مجلس الوزراء الإمضاء:سعد الدين الحريري
      </Signature>
  </Table>
  <Table1>
    <ArticleNumber>1</ArticleNumber>
    <ArticleTitle>عقد استثنائي</ArticleTitle>
    <ArticleContent>
      مادة ١ : يدعى مجلس النواب إلى عقد استثنائي يفتتح
      بتاريخ ٢٠١٨\٣\١٩ ويختتم بتاريخ ٢٠١٨\٩\٢٠
    </ArticleContent>
  </Table1>
  <Table1>
    <ArticleNumber>2</ArticleNumber>
    <ArticleTitle>برنامج اعمال</ArticleTitle>
    <ArticleContent>
      مادة ٢ : برنامج أعمال
      يحدد برنامج أعمال هذا العقد الاستثنائي بما يلي:
      -مشاريع موازنات الاعوام ٢٠١٦، ٢٠١٧ و ٢٠١٨
      -مشاريع القوانين والاقتراحات والنصوص
      التي يقرر مكتب المجلس طرحها على المجلس.
    </ArticleContent>
  </Table1>
  <Table1>
    <ArticleNumber>3</ArticleNumber>
    <ArticleTitle></ArticleTitle>
    <ArticleContent>
      ينشر هذا المرسوم ويبلغ حيث تدعو الحاجة
    </ArticleContent>
  </Table1>
</NewDataSet>
```

**Fig 5: Output XML Document**

## V.  FUTURE WORK

As further research, other types of juridical documents are to be investigated, mainly those that require error detection and correction prior to converting them into digital formats. Furthermore, other output data formats are to be studied, for instance JSON and YAML. Finally and in an independent endeavor, other automation operations are to be researched and developed basically transaction services in E-Governments.

## ACKNOWLEDGMENT

This research was funded by the Lebanese Association for Computational Sciences (LACSC), Beirut, Lebanon, under the "Arabic Programming Language Research Project – APLRP2019".

## REFERENCES

[1]   Georges Ifrah, "The Universal History of Computing: From the Abacus to the Quantum Computer", New York: John Wiley & Sons, ISBN 9780471396710, 2001

[2]   Padegs, A.,"System/360 and Beyond", IBM Journal of Research and Development IBM, vol. 25 no. 5, pp. 377–390, 1981

[3]   Gray, George T., Smith, Ronald Q., "Sperry Rand's Third-Generation Computers 1964-1980". IEEE Annals of the History of Computing, IEEE Computer Society, vol. 23 no. 1, pp.3–16, 2001

[4]   Amdahl, G. M., Blaauw, G. A., Brooks, F. P., "Architecture of the IBM System/360". IBM Journal of Research and Development, vol. 8 no. 2, pp.87–101, 1964

[5]   John L. Hennessy, David A. Patterson, David Goldberg, "Computer architecture: a quantitative approach", Morgan Kaufmann, ISBN 9781558607248, 2003

[6]   Fennell, Philip, "Extremes of XML", XML London Pub, ISBN 9780992647100, 2013

[7]   Kenneth C. Louden, "Compiler Construction: Principles and Practice", PWS Publishing Company, 1997, ISBN 0534939724

[8]   Hopcroft, John E., Motwani, Rajeev, Ullman, Jeffrey D., "Introduction to Automata Theory, Languages, and Computation (2 ed.)", Addison Wesley, 2001, ISBN 0201441241.

[9]   Chomsky, Noam, "Three models for the description of language", Information Theory, IEEE Transactions, vol. 2, no. 3, pp. 113–124, 1956.